\documentclass[preprint,12pt]{elsarticle}
\usepackage{graphicx, amssymb, amsfonts, amsmath, subfigure}
\journal{Physics Letters B}
\newcommand{\be}{\begin{equation}}
\newcommand{\ee}{\end{equation}}
\newcommand{\ba}{\begin{eqnarray}}
\newcommand{\ea}{\end{eqnarray}}
\DeclareMathOperator{\Tr}{Tr}

\begin{document}
\begin{frontmatter}
 \title {The bulk transition of QCD with twelve flavors and 
the role of improvement}
\author[Bern]{Albert Deuzeman}
\ead{deuzeman@itp.unibe.ch}
\author[Frascati]{Maria Paola Lombardo}
\ead{mariapaola.lombardo@lnf.infn.it}
\author[Groningen]{Tiago Nunes da Silva}
\ead{t.j.nunes@rug.nl}
\author[Groningen]{Elisabetta Pallante}
\ead{e.pallante@rug.nl}
\address[Bern]{ Albert Einstein Center for Fundamental Physics - University of Bern, Switzerland }
\address[Frascati]{INFN-Laboratori Nazionali di Frascati, I-00044, Frascati (RM), Italy}
\address[Groningen]{Centre for Theoretical Physics, University of Groningen, 9747 AG, Netherlands}

\begin{abstract}

We study the SU(3) gauge theory with $N_f=12$ flavors in the fundamental representation by use of lattice simulations with staggered fermions. 
With a non-improved action we observe  a  chiral  zero-temperature (bulk) 
transition separating a region at weak coupling, where chiral symmetry is realized, from a region at strong coupling  where chiral symmetry is broken. 
With improved actions, a more complicated pattern emerges, and in particular
two first order transitions in the chiral limit appear.  We observe that
at sufficiently strong coupling  the next-to-nearest neighbor terms of the improved lattice action are no longer irrelevant and can indeed modify the pattern observed without improvement. 
 Baryon number conservation can be realized
in an unusual way, allowing for an otherwise prohibited oscillating
term in the pseudoscalar channel.  We discuss the phenomenon  by means of explicit examples borrowed  from statistical mechanics. Finally, these observations can also be useful when simulating other strongly coupled systems on the lattice, such as graphene.
\end{abstract}


\end{frontmatter}

\section{Introduction}

In recent years attention has been drawn to the study of conformal symmetry restoration in non abelian gauge theories. On the one hand, there is theoretical interest in uncovering their phase diagram. On the other hand, the start of LHC activities creates the possibility of putting under scrutiny candidate scenarios for electroweak symmetry breaking, among others the possibility that strongly coupled dynamics govern the physics beyond the Standard Model.  Some of these models live in a quasi-conformal region of the parameter space at the TeV scale, such as walking technicolor or generalizations to composite Higgs models, or conformal symmetry might be thought to play a role at much higher energies.  
 
\begin{figure}[ht]
 \centering
\includegraphics[width=0.75\linewidth]{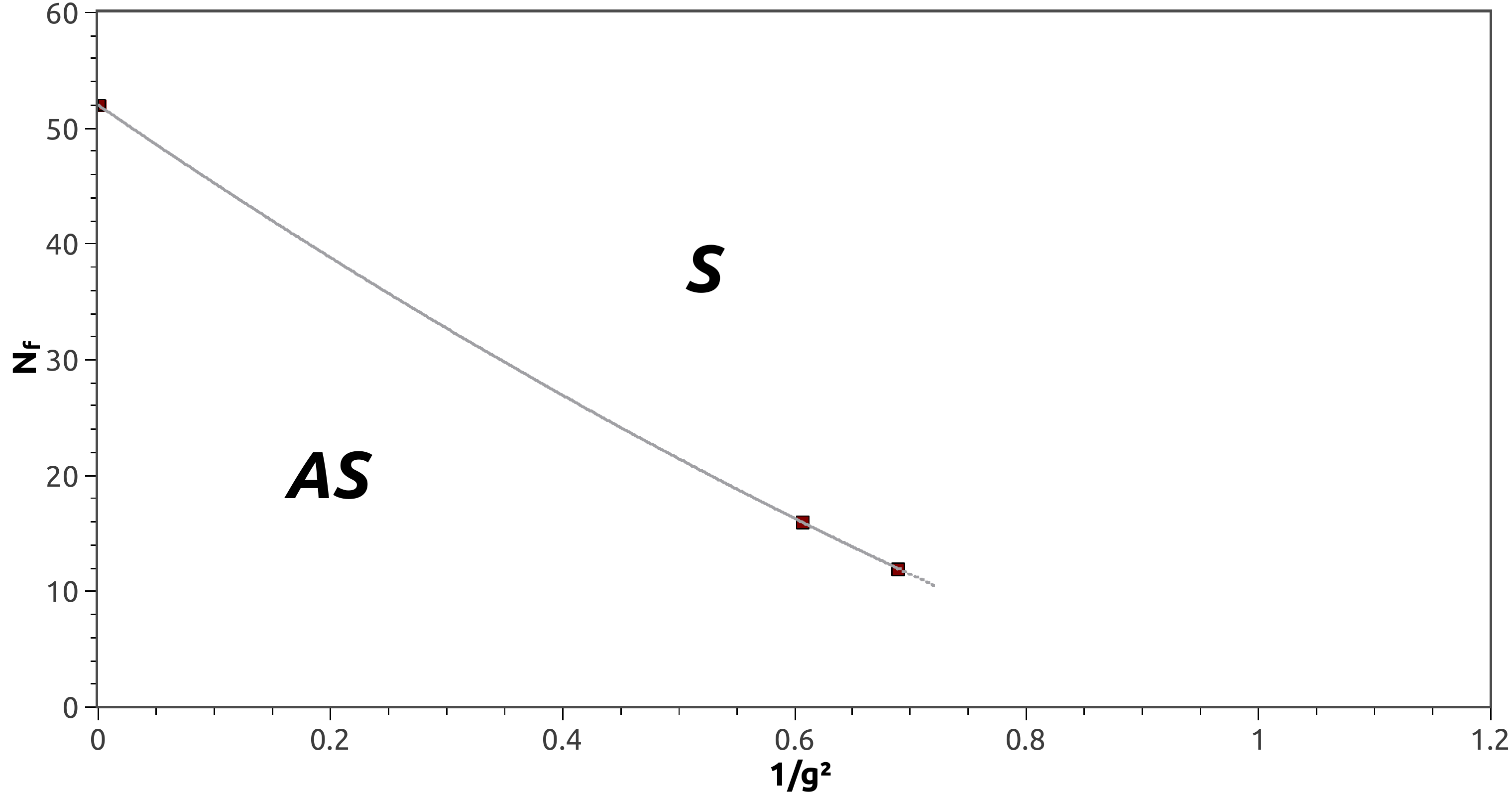}
 \caption{The bulk transition line in the $N_f$-$g^2$ plane of the phase diagram for SU(3) gauge theories with (unimproved) staggered fermions. 
The bulk transition separates a  QED-like, 
chirally symmetric, region (S, right side) from a chirally broken phase (AS, left side). 
The data points are all for a bare lattice fermion mass of 0.025 and should ideally be extrapolated to the chiral limit. 
Data for $N_f=16$ (work in progress) agree with Ref.~\cite{Damgaard1997169} after mass rescaling, the point at $g^2=\infty$ 
and $N_f \simeq 50$ is from Ref.~\cite{deForcrand:2012vh} and the point for $N_f=12$ is from the present work. The end point of the bulk line is unknown. Refs. \cite{Deuzeman2010a,Deuzeman:2011pa, Hasenfratz2011} reported a further bulk transition in the chirally symmetric phase. In this paper we argue that next-to-nearest neighbor interactions in the improved fermion action are necessary for the second transition to occur.}
 \label{fig:Miranskyplot}
\end{figure}

The main interest of these studies is of course the theory in the continuum limit. However, in recent years a growing amount of work has been devoted to the analysis of the so-called bulk transition emerging in the lattice phase diagram at strong bare gauge coupling, see Fig.~\ref{fig:Miranskyplot}. 
Early studies based on the strong coupling expansion of QCD predict that chiral symmetry is always broken in the
strong coupling limit, regardless of the number of flavors. Support for this claim was offered by Damgaard et al.~\cite{Damgaard1997169}, who uncovered a bulk transition for sixteen fundamental fermions. Later on, we have found \cite{DEUZEMAN2008} that QCD with eight fundamental fermions is still in the QCD phase, hence for $N_f=8$ the chirally restoring transition has a genuine thermal -- as opposed to bulk -- nature. These findings have been further confirmed by \cite{Jin2009}. A true bulk transition appears instead in QCD with twelve flavors, and the careful scrutiny of the region at its weak coupling side seemed consistent with exact chiral symmetry \cite{Deuzeman2010,Deuzeman:2011pa}. 
The presence of a zero temperature (bulk) transition between a chirally symmetric phase at weaker coupling and a chirally broken phase at stronger coupling is indicative of a theory being in the conformal window \cite{Deuzeman2010}. An analysis of the thermal transition in the preconformal region is again consistent with a critical number of flavors $N_f^c \simeq 12$ \cite{Miura:2011mc}. A review of current investigations of the $N_f=12$ theory, as well as a discussion of signatures and strategies can be found e.g. in 
\cite{Gies,Neil:2012cb,DelDebbio:2010zz,Pallante:2009hu}.

 Interestingly, and amusingly, a second bulk transition was uncovered by us \cite{Deuzeman2010a,Deuzeman:2011pa}, between the first observed bulk transition and the weak coupling region, where chiral symmetry studies were carried out. It is important to observe that, as all our analysis was done at the weak coupling side of such second transition, all our conclusions on the nature of the $N_f=12$ theory remain unaffected. Still, this was an interesting and unexpected observation calling for further analysis. The existence of a second transition was confirmed by the work in \cite{Hasenfratz2011,Schaich:2012fr}, where it was observed that the shift symmetry of staggered fermions was broken in the intermediate phase. Finally, a recent interesting development re-examined the early strong coupling studies: contrary to previous conclusions, it was observed that, with unimproved fermions, the line of bulk transitions ends for $N_f \simeq 51$. No second transition was observed in this case \cite{deForcrand:2012vh}. 

A more general line of work involving quantum -- or bulk -- transitions 
in a particle physics environment dates back to early studies of
QED at strong coupling. The transition in this context has been for a long time investigated in the hope
of finding an interacting, non asymptotically free theory in four dimensions. Such a theory requires a second order transition with non-trivial exponents. Indeed, the bulk transition for QCD with a large number of flavors has close similarities with the QED transition and in this spirit, inspired by the work in \cite{Kaplan2009}, we have proposed \cite{Deuzeman2010a}
to search for an interacting UVFP  at the bulk transition itself. One of us has  also explored this possibility in the context of AdS/CFT \cite{Barranco:2011vt}.
QED-like lattice systems are also being used for the simulation of strongly coupled graphene. Using an effective field theory description, 
the system can be modeled by QED in 2+1 dimensions,  whose bulk transition 
can be analyzed borrowing early lattice methods and strategies \cite{Drut:2012md}. 

In conclusion, bulk transitions are interesting for several 
reasons
ranging from a diagnostic of the conformal window to fundamental 
QFT questions and the physics of condensed matter systems, such as graphene. 
Lattice methods are mandatory for studying these phenomena. And it is important to realize that,
since we are not taking the continuum limit, lattice actions that are equivalent in the
continuum might have substantially different features at finite lattice spacing. A very well known example is offered by lattice 
QED in four dimensions, where the compact and non-compact formulation produce a different order for the phase transition. Less investigated, but for many reasons interesting, is the effect of improvement. 

This note is dedicated to the study of the bulk transition for $N_f=12$ and the role of improvement. 
As we have anticipated, the discovery of an intermediate 
phase at strong coupling with a peculiar behavior has recently attracted 
some interest. In this note we present our results on the nature of this phase, 
and we show that it only exists when the fermion sector is improved.

\section{The Actions}
\label{sec:setup}

We simulated the $SU(3)$ gauge theory with twelve  
flavors of staggered fermions in the fundamental representation.
In order to separate the effects of improvement for the gauge and fermion action,  we performed simulations for different cases, labeled A to D in Table~\ref{tab:actions}, with improvement present in the gauge and/or fermion sector.
\begin{table}
\begin{center}
\begin{tabular}{|c | c c |}
\hline
Action & Gauge Improvement & Fermion Improvement \\
\hline
 A  &No & No \\
B  & Yes & No \\
C  & No & Yes \\
D  & Yes&  Yes \\
\hline
\end{tabular}
\caption{Actions used in this work: gauge improvement refers to tree-level Symanzik improvement in the gauge action, while fermion improvement refers to tree-level Symanzik improvement of the staggered fermion action, i.e. the addition of the Naik term \cite{Naik:1986bn, Bernard:1997mz}.}
\label{tab:actions}
\end{center}
\end{table}
Many of the comparisons presented here are for a bare lattice fermion mass of 0.025 and a volume $16^3\times 24$, as in Fig.~\ref{fig:comp-imp}.
For some of the Actions we have explored an extended set of parameters, although
a complete presentation of our results will appear elsewhere.

\noindent Action A is given by:
\begin{equation}
S = -\frac{N_f}{4}{\Tr} \ln M(am,U) + \beta\, {\mbox{Re}}(1-U(\mathcal{P}))
\label{eq:naive_act}
\end{equation}
where $M(am,U)$ is the fermion matrix for the naive staggered action for a single flavor with mass $m$, $\beta =6/g^2$ is the SU(3) lattice coupling and $U(\mathcal{P})$ is the trace of the ordered product of link variables along the single plaquette $P$ divided by the number of colors. 

\noindent Tree-level Symanzik improvement of the gauge action leads to Action B, 
\begin{equation}
S = -\frac{N_f}{4}{\Tr} \ln M(am,U) + \sum_{i=0,1}\beta_i(g^2) \sum_{\mathcal{C}\in{\mathcal{S}_i}  }   {\mbox{Re}} (1-U(\mathcal{C}))
\label{eq:imp_act}
\end{equation}
where $U(\mathcal{C})$ are the traces of the ordered product of link variables along the closed paths $\mathcal{C}$ divided by the number of colors. 
The $\mathcal{S}_0$ contains all the $1\times 1$ plaquettes (nearest neighbors), while $\mathcal{S}_1$ contains all the $1\times 2$ and $2 \times 1$ rectangles (next-to-nearest neighbors). The couplings are defined as $\beta_0 = (5/3)\beta$ and $\beta_1 = -(1/12)\beta$, where $\beta = 6/g^2$ is the SU(3) lattice coupling of the unimproved gauge action. 

\noindent Improvement of the staggered fermion action is realized according to the Naik prescription \cite{Naik:1986bn, Bernard:1997mz}
\ba
S_F &&= a^4\sum_{x;\mu} \eta_\mu(x)\bar{\chi}(x)\frac{1}{2a}\left\{
c_1\left [ U_\mu(x)\chi(x+\mu ) -U^\dagger (x-\mu)\chi(x-\mu)   \right ] \right .\nonumber\\
&&+ c_2\left [ U_\mu (x)U_\mu (x+\mu)U_\mu(x+2\mu)\chi(x+3\mu)   \right . \nonumber\\
&&\left.\left . - U^\dagger_\mu(x-\mu)U^\dagger_\mu(x-2\mu)U^\dagger_\mu(x-3\mu)\chi(x-3\mu)\right ]
\right\} \nonumber\\
&&+a^4 m\sum_x\bar{\chi}(x)\chi(x)
\label{eq:naikaction}
\ea
where the phase factor $\eta_\mu(x) = (-1)^{(x_0+x_1\ldots +x_{\mu -1} )}$ and the action is written in terms of the one component staggered fermion fields $\chi(x)$. The coefficients $c_1=1$ and $c_2=0$ reproduce the naive staggered fermion action, while the Naik choice $c_1=9/8$ and $c_2=-1/24$ provides $O(a^2)$ accuracy at tree level. Notice that the additional Naik term involves up to third-nearest neighbor interactions. 

\section{Results}
\label{sec:results}
\begin{figure}[ht]
 \centering
\includegraphics[width=0.8\linewidth]{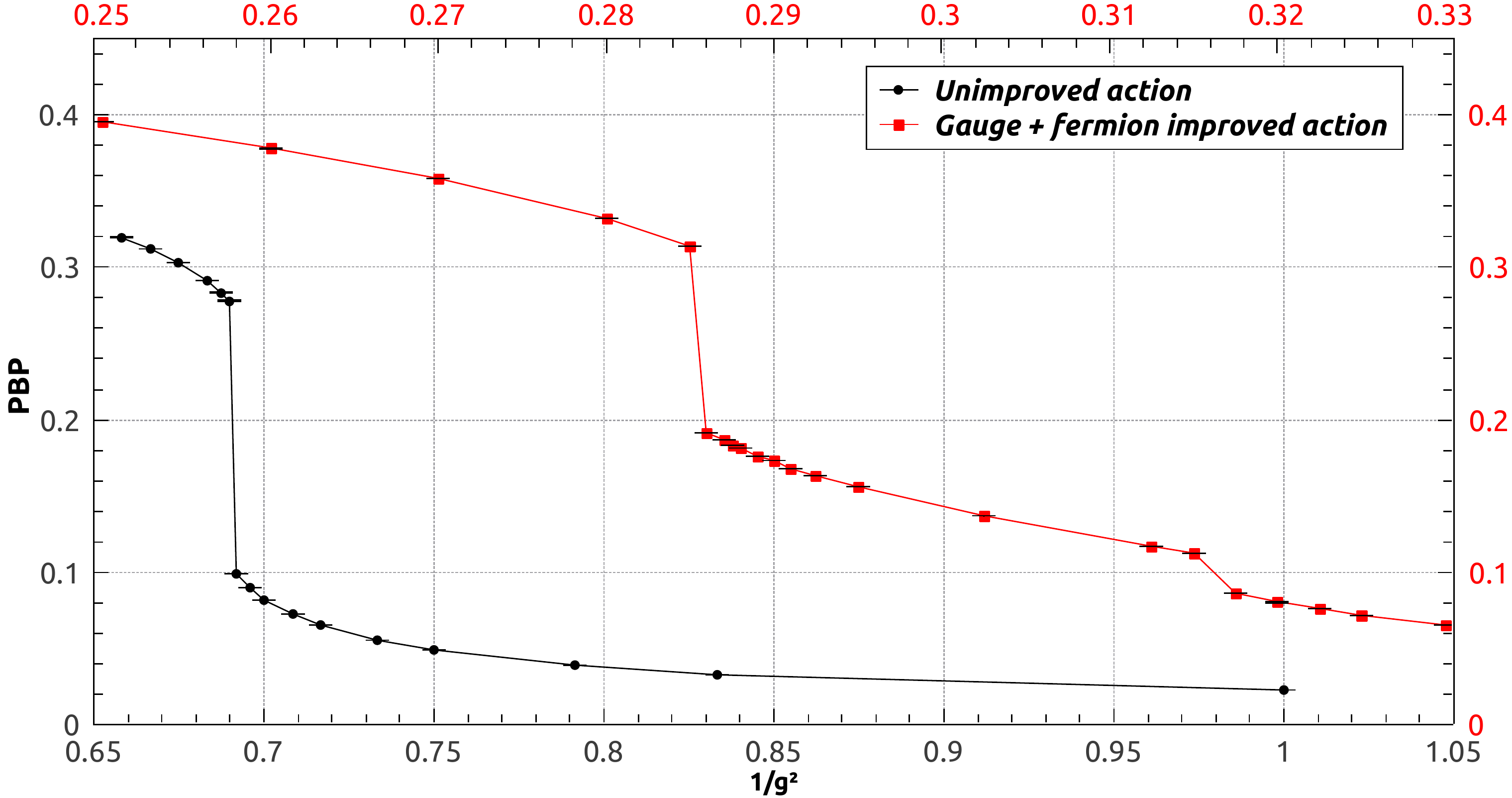}
 \caption{
The chiral condensate for the SU(3) gauge theory with $N_f=12$ fundamental flavors as a function of $1/g^2$, with $g$ the lattice bare coupling. We show the results for the unimproved action, Action A (leftmost, black) and for the improved gauge and fermion action, Action D (rightmost, red).
Data are for am=0.025 and volume $16^3\times 24$. The weaker coupling crossover of the improved action disappears in the unimproved case.}
\label{fig:comp-imp}
\end{figure}

The main numerical 
result of this work is summarized by Fig.~\ref{fig:comp-imp}, where the transition region for the fully improved and unimproved actions is shown. Two rapid crossovers are present with the improved action, while a single chiral symmetry breaking transition is present in the unimproved case.

\subsection{Action A: the unimproved case}
\begin{figure}
 \centering
 \subfigure[\label{fig:PBPnaive}]%
 {\includegraphics[width=0.49\columnwidth]{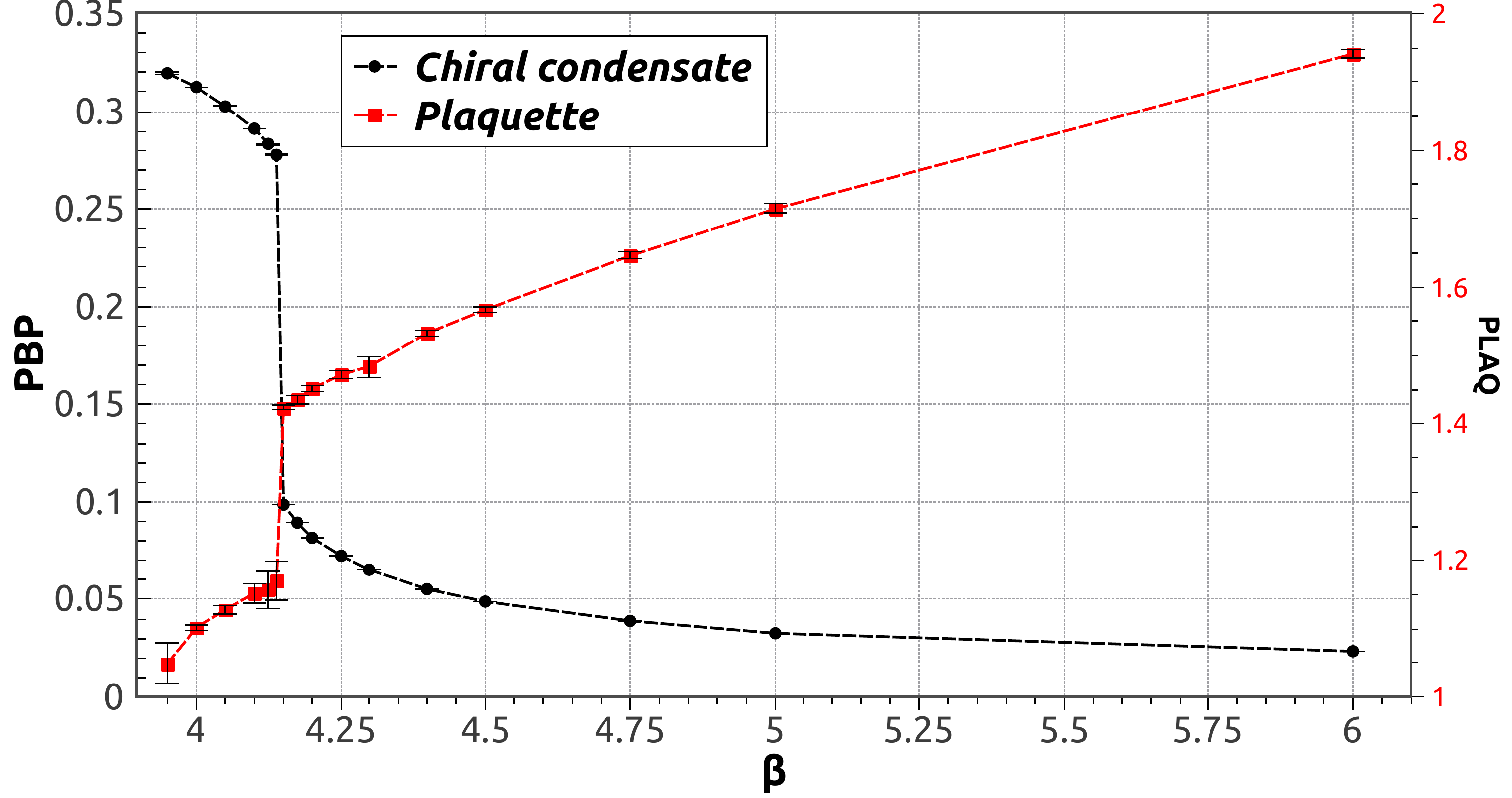}}
 \subfigure[\label{fig:Plaqnaive}]%
 {\includegraphics[width=0.49\columnwidth]{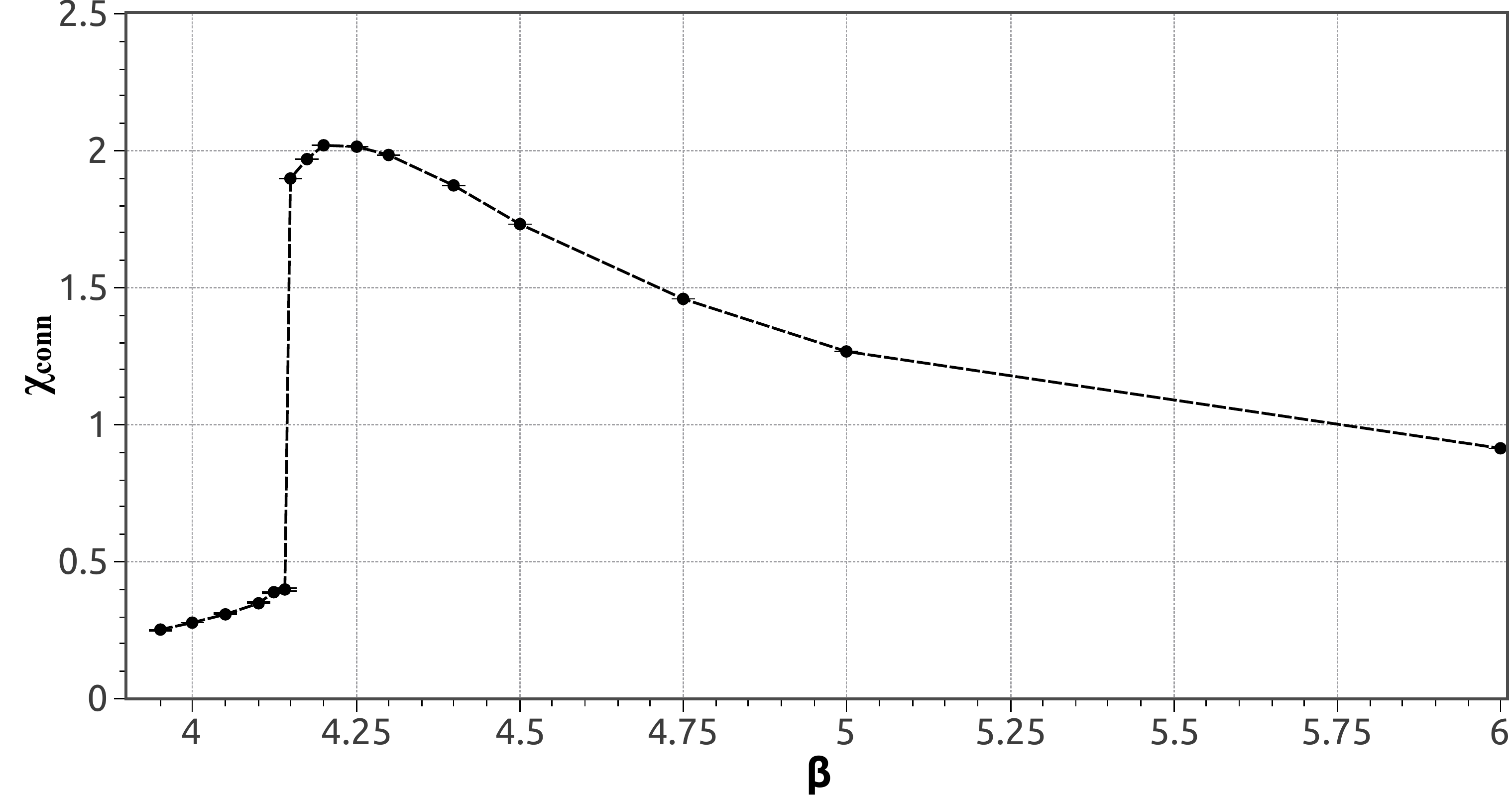}}
 \caption{(a) Rapid crossover of the chiral condensate (PBP) and the plaquette for $N_f=12$ flavors with the unimproved action (Action A) as a function of the lattice coupling $\beta = 6/g^2$  in the strong coupling region, for  $am=0.025$ and volume $16^3\times 24$. (b) The connected susceptibility for the same parameters.}
 \label{fig:PBPPlaq_Naive}
\end{figure}

Fig.~\ref{fig:PBPPlaq_Naive}(a) shows the 
 rapid crossover for the chiral condensate 
(left) superimposed on  the plaquette (right), at the bare lattice mass $am=0.025$. 
No additional structure is observed in the chiral condensate.
We corroborate these observations by showing 
the connected component of the chiral susceptibility $\chi_{conn}$ in 
Fig.~\ref{fig:PBPPlaq_Naive}(b); its behavior is as expected and no sign of 
 an intermediate phase at weaker coupling and additional transitions is present. 

\noindent Given the absence of phase transitions or indications for a  crossover,
it is plausible to conclude that  the weak coupling phase of this theory is  
continuously connected with the asymptotically free regime that admits a continuum limit\footnote{In other words, when no phase transition occurs at the infrared fixed point (IRFP) of the theory, the strong coupling QED-like side of the IRFP should be continuously connected to the asymptotically free weak coupling side.}. 
If this is true, its symmetry properties
are the same as the ones of the improved action, extensively investigated in our previous work. 
We then conclude that the rapid crossover observed for Action A
in Fig.~\ref{fig:PBPPlaq_Naive} should be interpreted
as the finite mass remnant of a
bulk chiral transition separating 
the chirally broken phase at strong coupling from the chirally symmetric 
phase, in complete analogy with the unimproved results of Ref.~\cite{Damgaard1997169}
for $N_f = 16$.
\subsection{Action D: the improved gauge and fermion action}
We now consider Action D, i.e. the case where both fermion and gauge actions are tree-level Symanzik improved. 
For small enough bare masses ($am \lesssim 0.04$), at the simulated volumes, two rapid crossovers are observed in the value of the chiral condensate: a large one at stronger coupling and a smaller one at weaker coupling (see Fig.~\ref{fig:comp-imp}). Preliminary results were reported in \cite{Deuzeman2010a, Deuzeman:2011pa}. 
As expected, the transition to the chirally broken phase moves towards stronger couplings when the action is improved. Less expected is the fact that the transition appears to be realized in two steps, leading to one intermediate region. 
%
\begin{figure}
 \centering
 \subfigure[\label{fig:massdep}]%
 {\includegraphics[width=0.49\columnwidth]{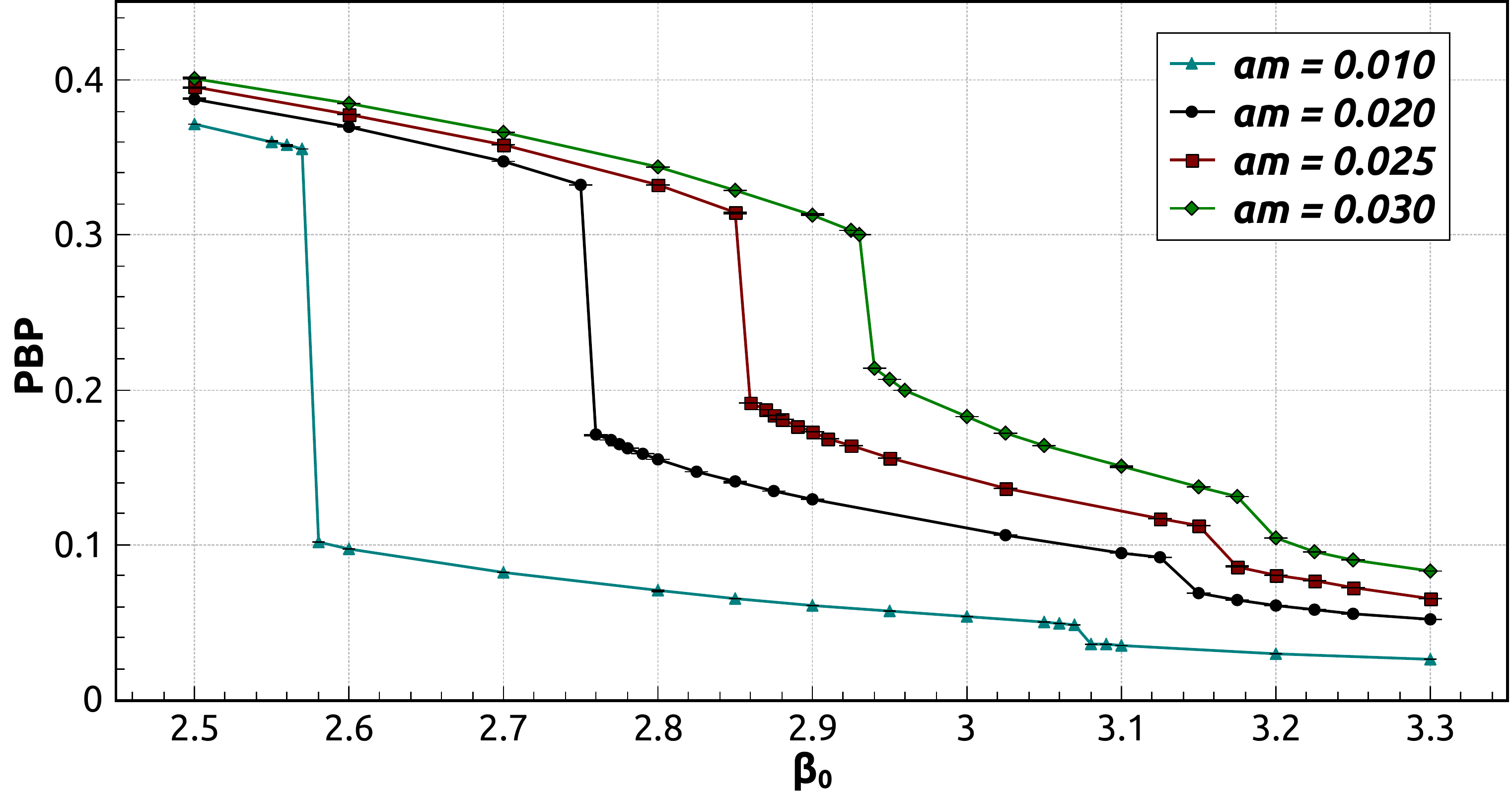}}
 \subfigure[\label{fig:ntdep}]%
 {\includegraphics[width=0.49\columnwidth]{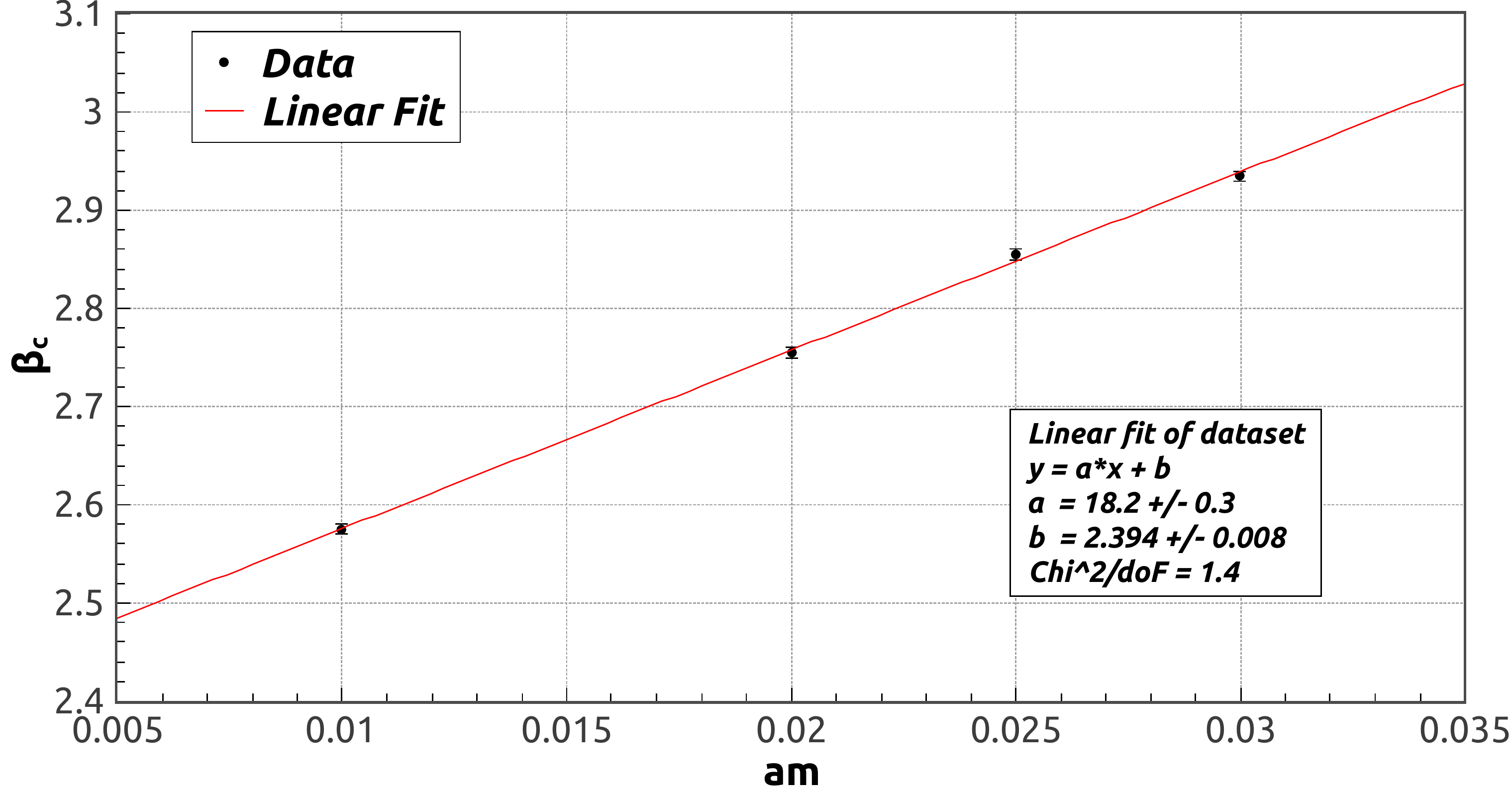}}
 \caption{(a) Rapid crossovers in the chiral condensate (PBP) with the improved action as a function of the coupling $\beta_0 = 10/g^2$  in the strong coupling region for different bare masses. (b) The mass dependence of the critical $\beta$ value extracted from the central point of the strong coupling is in agreement with a linear scaling expected for a first order transition.}
 \label{fig:first}
\end{figure}
Fig.~\ref{fig:first}(a) shows that the crossover at stronger coupling becomes more pronounced as the bare mass decreases. No dependence on the lattice temporal extent is observed and no perturbative scaling can be realized \cite{Deuzeman2010a, Deuzeman:2011pa}.  
The mass dependence of the location of the strong coupling rapid crossover in Fig.~\ref{fig:first}(b) is in agreement with a linear scaling expected for a first order transition. 

\begin{figure}
 \centering
 \subfigure[\label{fig:massdepconn}]%
 {\includegraphics[width=0.49\columnwidth]{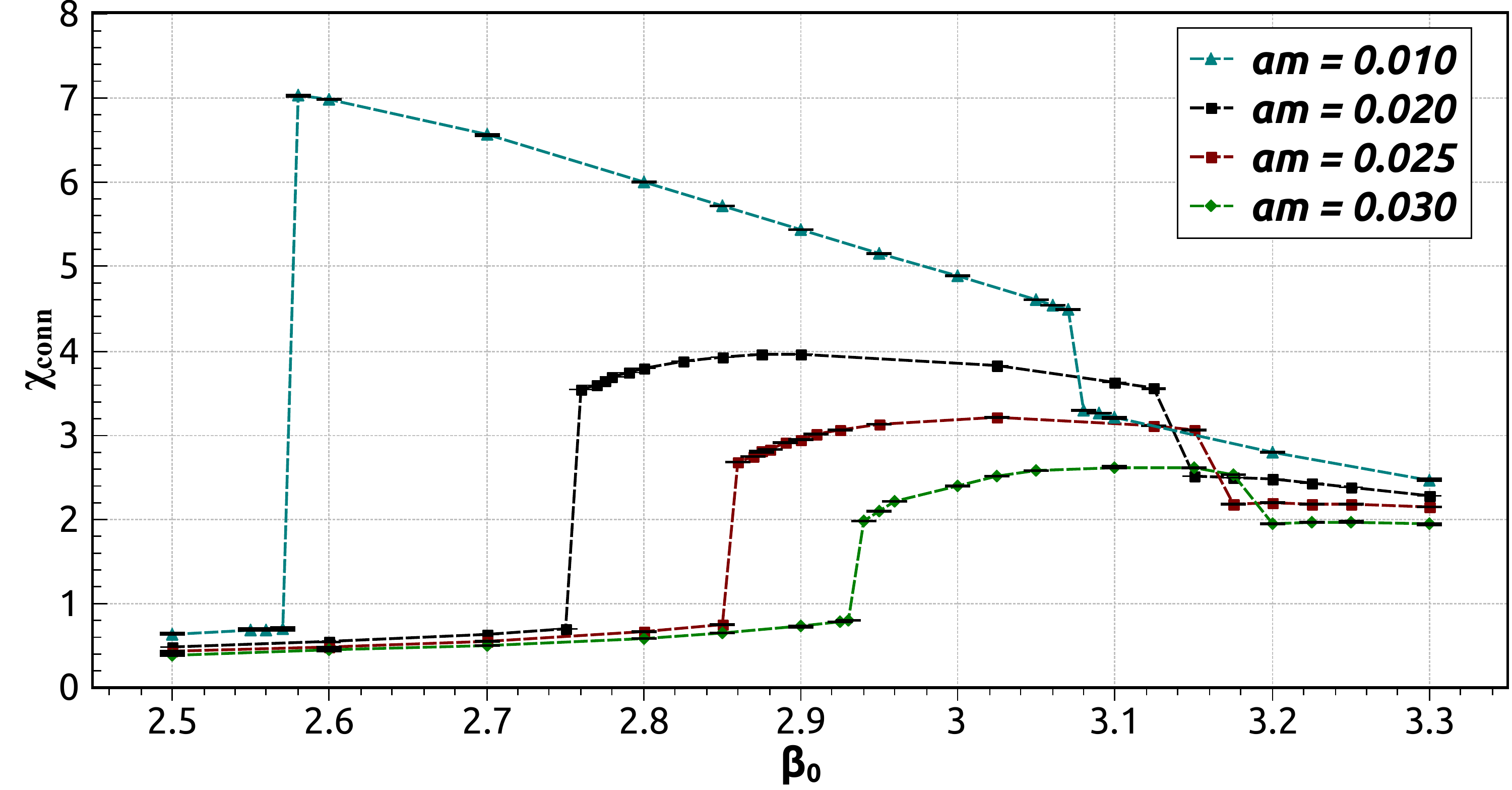}}
 \subfigure[\label{fig:massdepdisc}]%
 {\includegraphics[width=0.49\columnwidth]{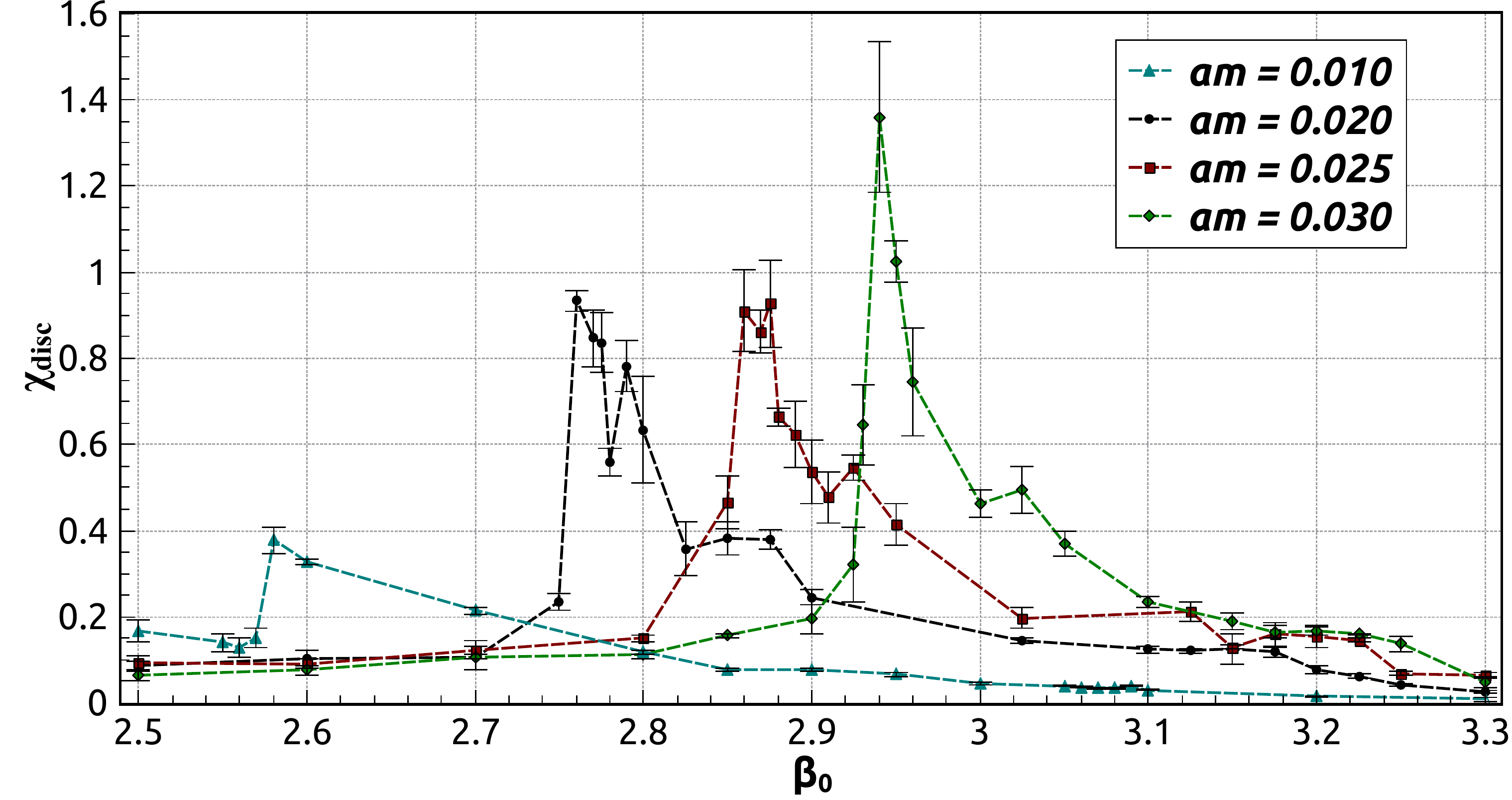}}
\caption{
Mass dependence of the connected chiral susceptibility (left) and the disconnected chiral susceptibility (right). }
 \label{fig:fourth}
\end{figure}

\noindent The disconnected component of the chiral susceptibility shows a pronounced peak only in correspondence to the strong coupling rapid crossover, as shown in 
Fig.~\ref{fig:fourth}(b). These results  indicate that the strong coupling rapid crossover is the one corresponding to chiral symmetry breaking.
\noindent Consider now the crossover in the chiral condensate at weaker coupling. The
hints at a jump become weaker as we approach the chiral limit, see Fig.~\ref{fig:first}(a).
On the other hand, the behavior of the chiral condensate as a function of the 
mass suggests a discontinuity in its mass derivative, which is best studied
by considering the chiral susceptibility.

\noindent The connected component of the chiral susceptibility exhibits near discontinuities at the condensate crossovers, as shown in Fig.~\ref{fig:fourth}(a). The magnitude of both discontinuities increases as the bare mass decreases. This suggests that the jump at weaker coupling also corresponds to
a genuine phase transition in the chiral limit, as suggested in \cite{Deuzeman2010a}
and confirmed in \cite{Deuzeman:2011pa,Hasenfratz2011}.  We conclude that we are observing two distinct phase transitions,
one associated with a change of the slope of the chiral condensate at weaker coupling, the other
with the chiral condensate itself at stronger coupling. 

\begin{figure}
 \centering
\includegraphics[width=0.6\linewidth]{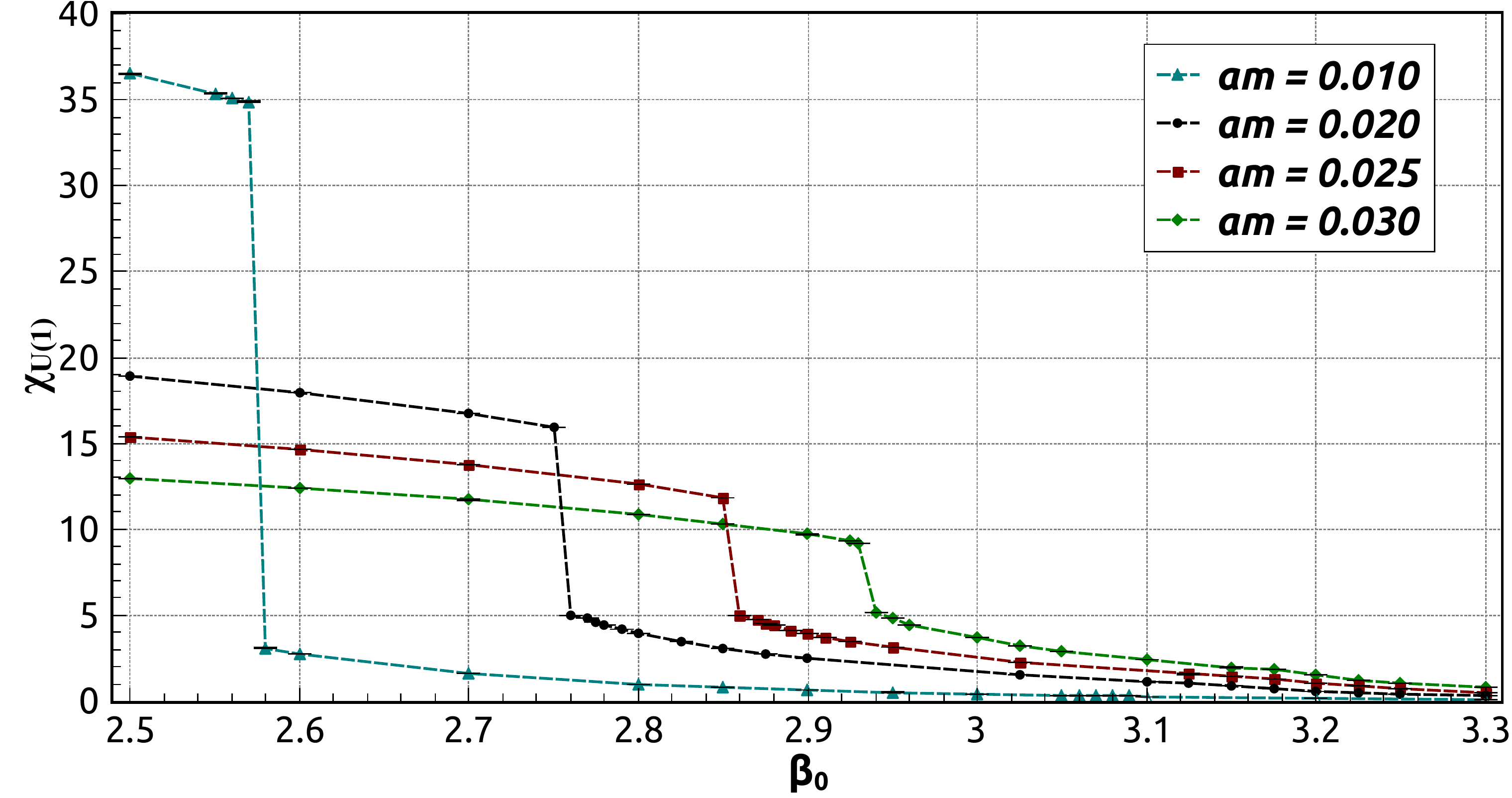}
 \caption{Mass dependence of the difference of the pseudoscalar susceptibility $\chi_\pi = \langle \bar{\psi}\psi\rangle /m$ and the connected scalar susceptibility. This order parameter probes $U_A(1)$ (effective) restoration in the continuum theory. Its behavior would suggest $U_A(1)$ restoration (see caveats in main text) at the chiral transition. }
\label{fig:UA1_cont}
\end{figure}

\noindent In the continuum language the observed pattern of the susceptibilities, and of the continuum order parameter shown in Fig.~\ref{fig:UA1_cont}, 
suggests $U_A(1)$ (effective) restoration at the strong coupling chiral transition.
However, a proper analysis of the axial anomaly 
is hampered at strong coupling by the absence of a conserved local flavor singlet current.
It would be interesting to further analyze the lattice non-local order parameter in the intermediate region in terms of the point-split staggered correlators, analogous to the finite temperature study in \cite{Kogut:1998rh}.

%
\subsection{The spectrum}
We recall that staggered meson correlators on a lattice with temporal extent $T$ and periodic boundary conditions have the general form
\be
\label{eq:stagcf}
C(t) = \sum_i\, A_i\left (e^{-m_i t} +e^{-m_i(T-t)}\right ) +(-1)^t \tilde{A}_i \left ( e^{-\tilde{m}_i t} + e^{-\tilde{m}_i(T-t)}\right ) \, .
\ee
For each state, the parity partner adds a component with alternating sign $(-1)^t$. This is a property  of the staggered formulation and it is true for all correlators with the exception of the equal mass Goldstone pseudoscalar correlator.
For equal quark and antiquark masses, the parity partner operator for the 
Goldstone pion is proportional to a charge density operator and thus its 
vacuum expectation value is zero.

\begin{figure}
 \centering
 \includegraphics[width=0.9999\linewidth]{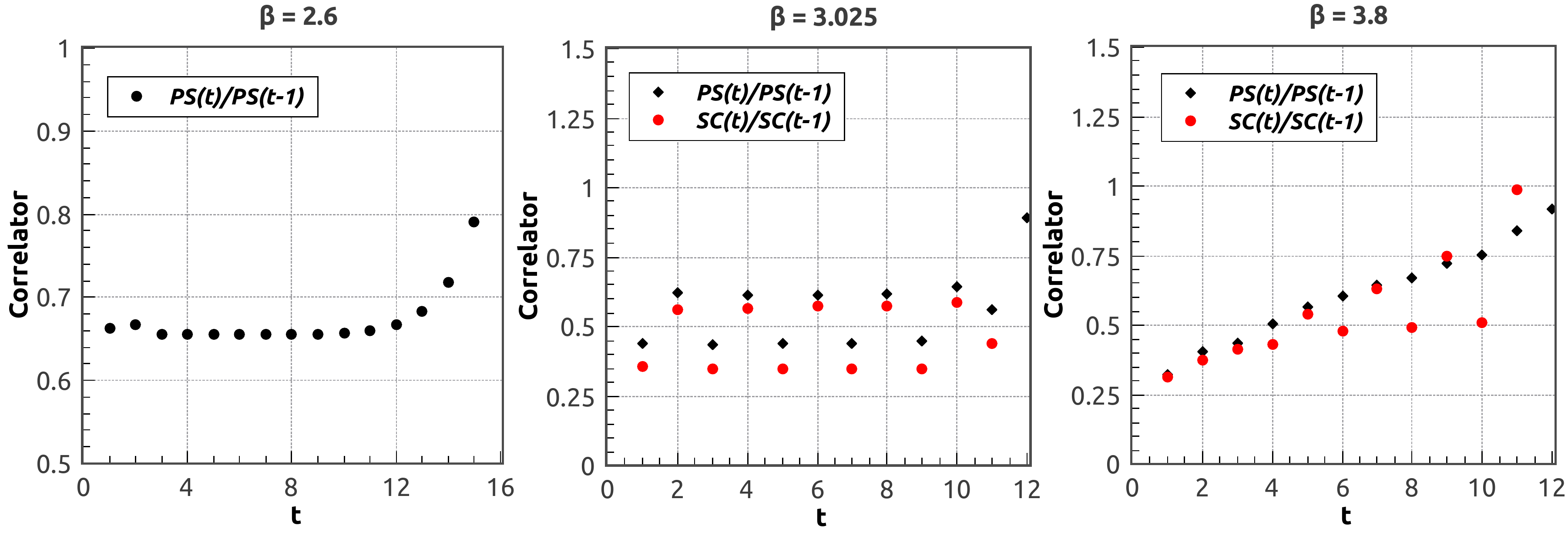}
 \caption{ Central values of the ratios $C(t)/C(t-1)$ for the pseudoscalar (PS, black circles) and scalar (SC, red squares) two-point correlation functions for coupling values in the three interesting regions. From left to right: the chirally broken phase, the intermediate phase and the weak coupling phase. The coupling in this case corresponds (left to right) to the improved $\beta_0=10/g^2 =$ 2.6, 3.025, 3.8.}
 \label{fig:oscillation}
\end{figure}

We give an overview of our results in Fig.~\ref{fig:oscillation}.
The most salient feature in Fig.~\ref{fig:oscillation} is an oscillatory component that arises for the pseudoscalar correlator in the intermediate region ($\beta_0=10/g^2 = 3.025$). This effect was also observed by the authors of \cite{Hasenfratz2011}. In this region chiral symmetry is exact and the scalar and pseudoscalar correlators should become increasingly degenerate by moving towards 
the chiral limit. 

\noindent What we see in Fig.~\ref{fig:oscillation}, moving from weak to strong coupling (right to left), is as follows. In the chirally symmetric region, the pseudoscalar and scalar correlators are close to each other. 
As expected, the staggered scalar correlator has an oscillating component while the pseudoscalar has not. The non-horizontal shape of the ratios indicates a significant contribution from excited states. 

 In the intermediate region, a new oscillating component arises in the pseudoscalar correlator, and seems to also arise in the scalar correlator for $\beta_0=3.025$. 
This is consistent with the abrupt change of slope in the mass dependence of the chiral condensate, given that the chiral susceptibility $\langle\bar{\psi}\psi\rangle /m$ equals the volume integral of the pseudoscalar correlator. 

 At strong coupling ($\beta_0 =2.6$) chiral symmetry is broken and the pseudoscalar lightest state is the Goldstone boson of the broken symmetry, thus very light and largely non degenerate with the scalar state. We observe that the oscillating component in the pseudoscalar correlator visibly decouples. 
\begin{figure}
 \centering
 \subfigure[\label{fig:asymmetry}]%
 {\includegraphics[width=0.75\columnwidth]{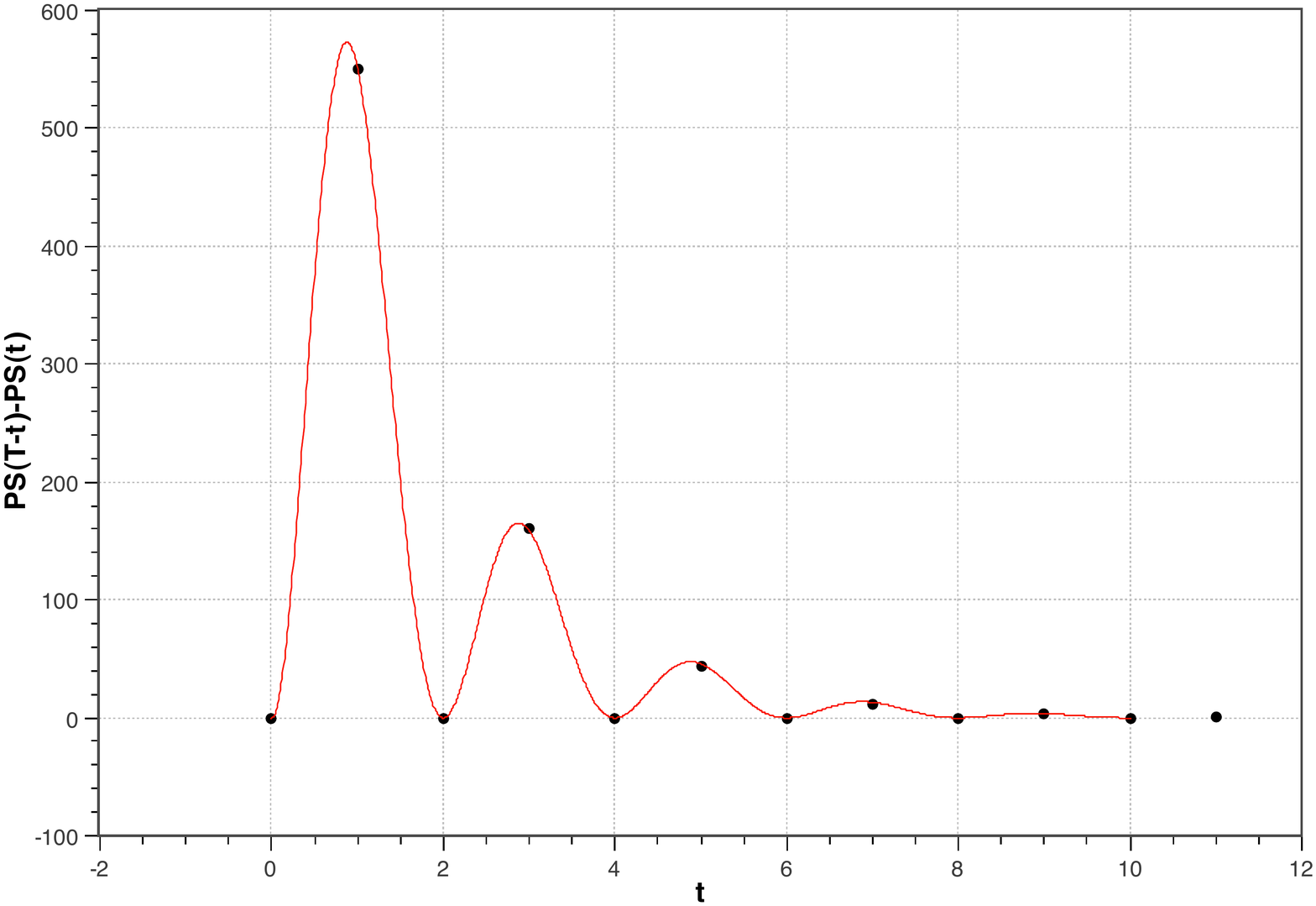}} 
 \caption{  The asymmetry of the Goldstone pseudoscalar correlator in the intermediate region ($\beta_0=3.025$), with superimposed the result of the fit to eq.~(\ref{eq:fit_asym}); the fitted parameters are $m\simeq 0.62$ and $C\simeq 1027$.  }
 \label{fig:deg_and_asym}
\end{figure}
A second observed effect is the presence of an asymmetry under $t\to T-t$ of all studied correlators in the intermediate region, i.e. $\beta_0 =3.025$. 
To highlight this asymmetry we have plotted the difference
$C(t) - C(T - t)$ for the pseudoscalar correlator in Fig.~\ref{fig:deg_and_asym}. We see that 
\ba
&&C(t) \neq C(T-t)~~~~ {\mbox{for t odd}}\nonumber\\
&&C(t)\sim C(T-t) ~~~~ {\mbox{for t even}}
\ea
In other words there is a violation of staggered-time reversal symmetry.
The asymmetry is well fitted, see  Fig.~\ref{fig:deg_and_asym}, by the functional form 
\be
C\left ( 1-(-1)^t \right ) \left ( e^{-mt} - e^{-m(T-t)}\right )
\label{eq:fit_asym}
\ee
with $C\simeq 1027$ and $m\simeq 0.62$ consistent with the fit of the pseudoscalar correlator on $t\ll T$. 

\noindent One caveat is in order: it is known that such an asymmetry may typically be present when configurations are not thermalized or statistics is too low.  For this reason we have increased thermalization time and statistics for this point  to a few times the ones in the other two regions. The asymmetry persists and does not vary with increasing thermalization or statistics. Hence, even if the observed asymmetric state is a metastable state, its tunneling probability to the opposite asymmetry seems to be extremely low, suggesting that a seed is indeed stabilizing it.

\subsubsection{Disentangling the effect of fermion and gauge improvement}
\begin{figure}
 \centering
 {\includegraphics[width=0.75\columnwidth]{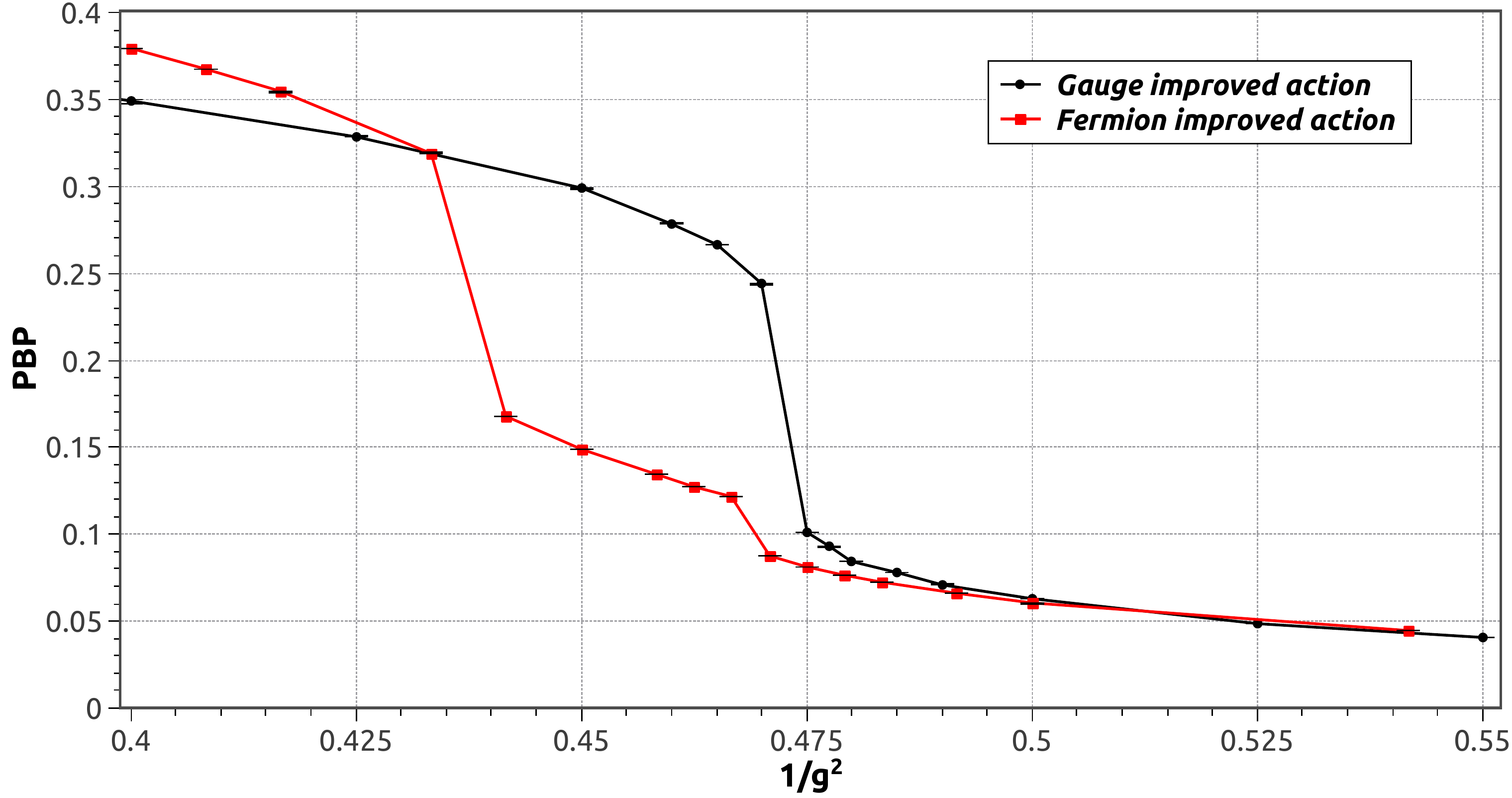}}
 \caption{ Two rapid crossovers are observed for the Naik improved fermion action (red), while one rapid crossover is observed for the Symanzik improved gauge action (black).}
 \label{fig:bulk_halfimp}
\end{figure}

In order to expose the separate effects of improvement of the fermion action and gauge action, we have performed two additional sets of lattice simulations with one improvement at a time - Action B and C in Table I.
 In Fig.~\ref{fig:bulk_halfimp} we show the results for the improved fermion action, or the improved gauge action. 
These results make clear that in the case under study the Naik improvement of the staggered fermion action is mainly responsible for the appearance of an intermediate region in the gauge coupling. 

\section{Discussion}
\label{sec:improvement}
The results presented above suggest that third-nearest neighbor terms
in the Naik improved fermion action are responsible for the appearance of an
intermediate phase.

 This is perhaps not unexpected. It is well known that models with competing 
interactions may give rise to non-homogeneous structures and  novel phase
transitions. One prototypical example is the axial next-to-nearest neighbor 
Ising model, known as the ANNNI model \cite{ANNNIReview}.
These effects have not been observed at weak coupling,
where non-nearest neighbor terms concur to a faster approach to the continuum
limit, but might well appear at strong coupling when those terms become 
relevant. It is quite possible that quantitative predictions for the
appearance and properties of the additional phase could be obtained in the framework
of a strong coupling expansion that takes 
the improvement term into account explicitly -- we do not pursue this here. 

\noindent Here, we provide a plausible argument that accounts
for the appearance of such an intermediate phase and its peculiar
properties: i) the emergence of an oscillating component of the staggered two-point correlation function in the pseudoscalar channel, and  
ii) the  asymmetry of all two-point correlation functions under $t\to T-t$. 

  The authors of \cite{Arisue1984} have considered the exactly solvable Ising chain (1D) of length $L$ with next-to-nearest neighbor interactions. This example is extremely instructive. There are two regions of parameters. In one region the eigenvalues of the transfer matrix are real and positive.
In the other region, pairs of complex conjugate eigenvalues appear.
Intuitively, the first region (region I) is where the nearest neighbor interaction is dominant, while the second region (region II) is where the next-to-nearest neighbor term becomes dominant. 

\noindent As observed in \cite{Arisue1984}, the two regions will also emerge in a Symanzik improved gauge action where the couplings $\beta_0$ and $\beta_1$ are fixed as a function of the inverse gauge coupling $\beta$. In other words, it is the competition of nearest neighbor and next-to-nearest neighbor interactions at increasingly coarse lattice spacing that causes the system to enter the second region.

The same argument can be repeated for the Naik improved staggered fermion action, with up to third-nearest neighbors. In this case, the emergence of complex eigenvalues of the transfer matrix can be understood by looking at the free lattice fermion propagator for a single flavor, given by
\be
S_F(p)^{-1} = \sum_\mu i \gamma_\mu \left (\frac{9}{8}\sin{p_\mu}-\frac{1}{24}\sin{3p_\mu}\right )
\ee
with $-\pi/2\leq p\leq \pi/2$. The interacting theory at strong coupling can in principle  significantly modify the coefficient of each sine contribution. In particular, the change of sign of the second term will induce a pair of imaginary poles (zero tri-momentum) in the massless dressed propagator, i.e. ghosts will appear\footnote{It is known that the dispersion relation for Naik improved staggered fermions always contains complex roots at non-zero tri-momentum. All ghosts generated by the improvement decouple in asymptotically free theories when approaching the continuum limit.}. This signals the emergence of region II, likely the intermediate phase we have observed. 

\noindent
It would certainly be interesting to understand more quantitatively the connection between the poles in the quark propagator -- as emerging from the non-Hermiticity 
of the transfer matrix with Symanzik improvement \cite{Luscher1984} -- and the detailed structure of the two-point correlation functions in the intermediate phase.
We postpone this analysis hopefully to future work.
Here, we offer a qualitative explanation as to why a chirally symmetric phase
with the observed exotic features can appear 
in a gauge theory with fermion improvement. 

 In general, the occurrence of an oscillatory secondary state in the pseudoscalar (Goldstone) correlator with staggered fermions is forbidden by the baryon current conservation.
With improvement of the action, the total fermionic current will include additional terms 
which in turn define a modified form of the baryon number operator at zero chemical potential.
For the Naik improved free fermion action this
construction has been explicitly given by Gavai \cite{Gavai:2002uj}.
In the interacting case, a simple construction that should suffice for our purpose starts
with implementing 
 the  Kogut-Hasenfratz-Karsch prescription \cite{Kogut:1983ia, Hasenfratz:1983ba}
$U(x) \to \exp(\mu) U(x)$, $U^\dagger(x) \to \exp(-\mu) U^\dagger(x)$ along the temporal direction.  The total baryon number density is then 
\begin{equation}
 n (\mu) = d/d_\mu log Z(\mu)  =  n_1 (\mu)  + n_3(\mu)
\end{equation}
where $n_1 (\mu)$ comes from local interactions and $n_3 (\mu)$ 
comes from the third-nearest neighbor term. 
At vanishing chemical potential
the total density $n(\mu=0)$ must vanish due to baryon number conservation. 
This can be realized in two ways, either $n_1(\mu=0) = n_3(\mu=0) = 0$, or 
$n_1(\mu=0) = -n_3(\mu=0) \ne 0$.
When the vanishing baryon number is realized in the second way, a non-zero oscillating component is allowed to appear in the (Goldstone) pseudoscalar channel, as its coefficient is roughly speaking, proportional to $n_1$. At the same time, $n_1$ is also a 
measure of the forward-backward asymmetry. Hence, $n_1 \ne 0$ allows an oscillating term
 in the pseudoscalar channel and a time asymmetry in all correlators; this is indeed what we observe for the pseudoscalar correlator and the other correlators in the intermediate region. 

 Putting all the elements together, we would then arrive at this simplified picture: with an improved staggered fermion action
the occurrence of imaginary poles of the quark propagator (or equivalently complex eigenvalues of the transfer matrix) opens the possibility of intermediate phases. 
In the chirally symmetric phase, and for sufficiently weak coupling,
the tendency towards degeneracy of the scalar and pseudoscalar propagators
is contrasted by the requirement of zero baryon number; the latter forces
 the amplitude of the oscillating component in the pseudoscalar channel to vanish
($0=n \simeq n_1$), while the oscillating component in the scalar channel starts appearing.  
Towards stronger coupling, third-nearest neighbor interactions in the improved fermion action become increasingly relevant. Now, chiral symmetry (i.e. the degeneracy between scalar and pseudoscalar correlators)  can still be preserved by allowing $n_3 = - n_1 \ne 0$. In this way the oscillating component appears in the pseudoscalar channel and the temporal asymmetry appears in all channels. 
This is the intermediate phase. When the coupling grows even larger,
chiral symmetry is finally broken, the lightest pseudoscalar is its Goldstone boson, and scalar and pseudoscalar correlators can depart from each other.  Our observations in the broken phase (see Fig.~\ref{fig:oscillation}) are consistent with a situation where the conservation of baryon number is again realized in the usual way.  

 We note that such a characterization of the intermediate phase is consistent with 
the breaking of the shift symmetry discussed in \cite{Hasenfratz2011}. In fact, the `partial baryon number operators'
$n_1$ and $n_3$  might well be directly related to the operators measuring the
breaking of the shift symmetry. In addition, the presence of ghost poles in the quark propagator of the improved fermion action can be translated into the presence of complex eigenvalues of the improved transfer matrix, and in the case of staggered fermions the (real time) transfer matrix is related to the shift operator as $T=S_4^2$.  

\vspace{0.8truecm}

Our study shows that the emergence of an exotic intermediate phase in the chirally symmetric SU(3) gauge theory with twelve fundamental (staggered) flavors is due to the improvement of the fermion action, which adds next-to-nearest neighbor interactions that compete with the local terms at strong coupling. 

 These observations might be of interest to model builders, when needing to 
realize exotic intermediate structures in interacting gauge models with a relatively simple and controlled procedure. It is also amusing to notice that it is possible to
mimic features of a dense system and a complex action (time asymmetry) 
by working with a real action, without the sign problem.
Of course, from the perspective of the study  of the phase diagram for SU(N) gauge 
theories with many flavors, the observed features remain a peculiar form of lattice artifacts that should be well disentangled from the underlying physics of the system.

\section*{Acknowledgments}

This work was in part based on the MILC collaboration's public lattice gauge theory code. Computer time was provided through the Dutch National Computing Foundation (NCF) and the University of Groningen. MpL wishes to acknowledge the hospilatity of the GGI, program Lattice Field Theory, during the completion of this work.

\bibliographystyle{elsarticle-num}
\bibliography{ua1refs.bib}

\end{document}